\documentclass[10pt,a4paper]{article}
\usepackage{graphics,graphicx,array,hhline,color,float,mdwlist,amsmath,amsfonts,amssymb,cite,multirow,xcolor,subcaption,hyperref,tabularx,placeins,xspace}
\usepackage[left=2cm,right=2cm,top=2cm,bottom=2cm,head=1cm,foot=1cm]{geometry}

\usepackage{t1enc}
\usepackage[utf8]{inputenc}

\begin{document}
\title{A Particle Emitting Source From an Accelerating, Perturbative Solution of Relativistic Hydrodynamics}
\author{B\'alint Kurgyis$^{1}$,
M\'at\'e Csan\'ad$^{1}$\\
$^1$ E\"otv\"os Lor\'and University, H-1117 Budapest, P\'azm\'any P. s. 1/A, Hungary}
\maketitle

\begin{abstract}
The quark gluon plasma is formed in heavy-ion collisions, and it can be described by solutions of relativistic hydrodynamics. In this paper we utilize perturbative hydrodynamics, where we study first order perturbations on top of a known solution. We investigate the perturbations on top of the Hubble flow. From this perturbative solution we can give the form of the particle emitting source and calculate observables of heavy-ion collisions. We describe the source function and the single-particle momentum spectra for a spherically symmetric solution.
\end{abstract}

\section{Introduction}
 
Our aim is to study the role of acceleration in heavy-ion collisions under an analytic framework. There are many numerical simulations to solve the equations of relativistic hydrodynamics. However, the analytic solutions are also important in understanding the connection between the initial and final state of the matter. The equations of relativistic hydrodynamics can be treated perturbatively to generalize an already known exact solution. 
We will utilize the known solution Hubble-flow  \cite{Csorgo:2003ry} and a perturbative solution, which includes a pressure gradient and acceleration as perturbations on top of the original solution and was given in \cite{Kurgyis:2017zxg}. From this perturbative solution we can calculate the source function and study the role of the parameters and compare the observables to the ones calculated from the exact solution \cite{Csanad:2009wc}. 

\section{General Equations}
We are using the equations of relativistic perfect fluid hydrodynamics. This can be formulated as the following:

\begin{equation}
    \partial_\mu T^{\mu\nu}=0,
\end{equation}
where $T^{\mu\nu}$ is the energy-momentum tensor, which can be expressed with the four-velocity $u^\mu$, pressure $p$ and energy density $\epsilon$; and is the following for perfect fluids:

\begin{equation}
    T^{\mu\nu}=(\epsilon+p)u^\mu u^\nu - pg^{\mu\nu}.
\end{equation}

We denote the Minkowskian metric tensor by $g^{\mu\nu}=\text{diag}(1,-1,-1,-1)$, and we use $c=1$ notation.
In addition, we use a simple equation of state (EoS), where energy density is proportional to pressure, and $\kappa$ is constant:

\begin{equation}
    \epsilon=\kappa p.
\end{equation}

With this EoS the equations of relativistic hydrodynamics can be separated into the following Euler equation and energy equation:

\begin{align}
    \kappa u^\mu\partial_\mu p+(\kappa+1)p\partial_\mu u^\mu&=0,\\
    (\kappa+1)pu^\mu\partial_\mu u^\nu&=(g^{\mu\nu}-u^\mu u^\nu)\partial_\mu p.
\end{align}

Finally, we assume that there is a conserved charge density ($n$), therefore we can formulate a continuity equation for this conserved quantity:

\begin{equation}
    \partial_\mu (u^\mu n)=0.
\end{equation}
\section{Hubble-Flow and Its Perturbations}
There are several analytic solutions for the equations of relativistic hydrodynamics. In this paper we investigate the perturbations on top of the Hubble flow.
\subsection{Hubble-Flow}
The relativistic Hubble-flow is a 1+3D solution without acceleration or 
 pressure gradient \cite{Csorgo:2003ry}. It describes a self-similar expansion. The solution has the following form:

\begin{align}\label{u}\
u^\mu&=\frac{x^\mu}{\tau},\\ \label{n}
n&=n_0\left(\frac{\tau_0}{\tau}\right)^3\mathcal{N}(S),\\ 	\label{p}
p&=p_0 \left(\frac{\tau_0}{\tau}\right)^{3+\frac{3}{\kappa}}.
\end{align}

Here we denote the proper time by $\tau=\sqrt{x_\mu x^\mu}$. The self-similarity of the solution is ensured through the scale parameter $S$:

\begin{equation}
    u_\mu\partial^\mu S=0.
\end{equation}
\subsection{Perturbations on Top of the Hubble-Flow}
There are different generalizations of the above mentioned Hubble-flow \cite{Csanad:2014dpa,Csanad:2012hr}. Next, we would like to include acceleration and a pressure gradient as perturbations. A set of solutions for the first order perturbations on top of the original solution was given in \cite{Kurgyis:2017zxg}:

\begin{align} \label{du01}
\delta u^\mu&=\delta \cdot F(\tau) g(x_\mu)\partial^\mu S\chi (S),\\ \label{dp01}
\delta p&=\delta\cdot p_0\left(\frac{\tau_0}{\tau}\right)^{3+\frac{3}{\kappa}}\pi (S),\\ \label{dn01}
\delta n&=\delta \cdot n_0\left(\frac{\tau_0}{\tau}\right)^3 h(x_\mu)\nu (S).
\end{align}

This is a solution if the following conditions for the functions of the scale parameter and the newly introduced $h,F,g$ functions are satisfied:

\begin{align}\label{chi:1}
\frac{\chi'(S)}{\chi(S)}&=-\frac{\partial_\mu\partial^\mu S}{\partial_\mu S\partial^\mu S}-\frac{\partial_\mu S \partial^\mu \ln g(x_\mu)}{\partial_\mu S\partial^\mu S},\\ \label{pi:1}
\frac{\pi'(S)}{\chi(S)}&=(\kappa+1)\left[F(\tau)\left(u^\mu\partial_\mu g-\frac{3g(x_\mu)}{\kappa\tau}\right)+F'(\tau)g(x_\mu)\right],\\ \label{nu:1}
\frac{\nu (S)}{\chi(S)\mathcal{N}'(S)} &=-\frac{F(\tau)g(x_\mu) \partial_\mu S\partial^\mu S}{u^\mu\partial_\mu h(x_\mu)}.
\end{align}
\subsection{A Concrete Solution}\label{subsect:rjtj}
For further studies we chose a simple solution, which is more general than that was investigated in \cite{Kurgyis:2018sgx}. The scale parameter in this case is:

\begin{equation}
    S={r^j}/{t^j}.
\end{equation}

The perturbations are the following:

\begin{align}\label{du:rjtj}
\delta u^\mu&=\delta \cdot \left(\tau+a\tau_0\left(\frac{\tau}{\tau_0}\right)^\frac{3}{\kappa}\right)S^{-\frac{j+1}{j}} \partial^\mu S,\\ \label{dp:rjtj}
\delta p&=\delta\cdot p_0\left(\frac{\tau_0}{\tau}\right)^{3+\frac{3}{\kappa}}\frac{(\kappa+1)(\kappa-3)}{\kappa}j  S^{-\frac{1}{j}},\\ \label{dn:rjtj}
\delta n&=\delta \cdot n_0\left(\frac{\tau_0}{\tau}\right)^3 \left(\ln\left(\frac{\tau}{\tau_0}\right)+ a\frac{\kappa}{3-\kappa}\left(\frac{\tau}{\tau_0}\right)^{\frac{3}{\kappa}-1}\right)j^2  S^{\frac{j-1}{j}}\left(S^\frac{2}{j}-1\right)\left(1-S^{-\frac{2}{j}}\right)\mathcal{N}'(S).
\end{align}

For the scale function of the original charge density we chose a Gaussian shape:

\begin{equation}
    \mathcal{N}(S)=e^{-\frac{b r^2}{\dot{R_0}^2t^2}}=e^{-\frac{b}{\dot{R_0^2}}S^{2/j}}.
\end{equation}

This solution has the free parameters $\tau_0, n_0, p_0, \kappa$ and $b$ which are the same as in the original Hubble-flow. In addition to this, for the perturbations there are three new parameters: the perturbation parameter $\delta$, a dimensionless parameter $a$ and the exponent of scale parameter $j$. 
\section{Calculation of Observables}
In heavy-ion collisions, the velocity field, pressure and energy density can not be measured directly. Let us now investigate the quantities that can be measured in heavy-ion collisions and calculated from hydrodynamical solutions. For this we assume that the particles come from a thermalized medium of quark-gluon plasma and this can be characterized by a source which comes from a relativistic Jüttner-distribution similarly as in \cite{Csanad:2009wc, Kurgyis:2018sgx}. Also, we assume a constant freeze-out hypersurface in proper time at $\tau_0$. The temperature of the system is defined through the following equation: $p=nT$. For the pertubative handling we will have to calculate the first order perturbation of this source function. For the most general set of perturbations of the Hubble-flow described in Equations \ref{du01} and \ref{dn01} the source function has the following form:

\begin{align*}
S(x,p)=&N\delta(\tau-\tau_0)\mathrm{d}\tau \mathrm{d}^3 x n_0 \left(\frac{\tau_0}{\tau}\right)^3 \mathcal{N}(S) \exp\left[-\frac{p_\mu u^\mu}{ T_0\left(\frac{\tau_0}{\tau}\right)^{\frac{3}{\kappa}}}\mathcal{N}(S)\right]\left(\frac{\tau p_\mu u^\mu}{t}\right)\cdot\\
&\cdot\Bigg[1+\delta\Bigg(-\frac{F(\tau)g(x_\mu)\partial^0 S\chi(S)\tau}{t}+\frac{F(\tau)g(x_\mu)\chi(S)t}{\tau p_\mu u^\mu}p_\mu \partial^\mu S+\frac{F(\tau)g(x_\mu)\chi(S)}{T_0\left(\frac{\tau_0}{\tau}\right)^{\frac{3}{\kappa}}}\\
&+\frac{(p_\mu u^\mu)(\mathcal{N}(S)\pi(S)-h(x_\mu)\nu(S))}{ T_0 \left(\frac{\tau_0}{\tau}\right)^\frac{3}{\kappa}}+\frac{h(x_\mu)\nu(S)}{\mathcal{N}(S)}\Bigg)\Bigg]_,
\end{align*}
where $N$ is a normalization factor, and $p_\mu$ is the four-momentum of the outgoing particles. For further studies we use a Gaussian approximation of the source. This means that we write the source as the product of a Gaussian peak and some other terms. By performing the proper time integral we can study the spatial dependence of the source. 
In the case of the concrete solution described in Subsection \ref{subsect:rjtj} the source becomes a two component Gaussian:

\begin{align}
S(x,p) \mathrm{d}^3 x&=I_1+I_2\text{, where}\\
I_1&= N n_0 \zeta^{(1)} f_0 \left(1+\epsilon_1 +\epsilon_2+\epsilon_3\right) \mathrm{d}^3x, \\
I_2&=N n_0 \zeta^{(2)} f_0 \left(\epsilon_4+\epsilon_5\right) \mathrm{d}^3x.
\end{align}

With $\epsilon_i$ corresponding to the perturbative terms: 

\begin{align}\label{eq:e1}
    \epsilon_1 &= \delta j\frac{2ab\kappa\tau_0^4}{(\kappa-3)\dot{R_0}^2r(\tau_0^2+r^2)^{3/2}},\\
    \epsilon_2 &= \delta j\frac{(1+a)\tau_0^2}{r(\tau_0^2+r^2)^{1/2}},\\
    \epsilon_3 &= \delta j\frac{(1+a)\tau_0^2\left((p_xx+p_yy+p_zz)(\tau_0^2+r^2)^{1/2}-r^2E\right)}{r^3\left(E(\tau_0^2+r^2)^{1/2}-p_xx-p_yy-p_zz\right)},\\
    \epsilon_4 &= \delta j\frac{(1+a)\tau_0\left(r^2E-(p_xx+p_yy+p_zz)(\tau_0^2+r^2)^{1/2}\right)}{T_0r^3},\\ \label{eq:e5}
    \epsilon_5 &= \delta j\frac{\left(E(\tau_0^2+r^2)^{1/2}-p_xx-p_yy-p_zz\right)(2ab\kappa^2\tau_0^2+\dot{R_0}^2(3-\kappa)^2(\kappa+1)(\tau_0^2+r^2)^{2})}{\tau_0T_0\dot{R_0}^2\kappa(3-\kappa)r(\tau_0^2+r^2)^{3/2}},
\end{align}
with $r$ being the radial distance $r=\sqrt{x^2+y^2+z^2}$ and $f_0$ being the following function: 

\begin{equation}
    f_0=\frac{E\sqrt{\tau_0^2+r^2}-p_xx-p_yy-p_zz}{\sqrt{\tau_0^2+r^2}}.
\end{equation}

The $\zeta^{(1)}$, $\zeta^{(2)}$ have the following form in the Gaussian-approximation:

\begin{align}
\zeta^{(1)}&=\exp\left[-\frac{E^2+m^2}{2ET_0}-\frac{p^2}{2ET_\text{eff}}\right]\exp\left[-\frac{\left(x-x_s^{(1)}\right)^2}{2R^2}-\frac{\left(y-y_s^{(1)}\right)^2}{2R^2}-\frac{\left(z-z_s^{(1)}\right)^2}{2R^2}\right],\\
\zeta^{(2)}&=\exp\left[-\frac{E^2+m^2}{2ET_0}-\frac{p^2}{2ET_\delta}\right]\exp\left[-\frac{\left(x-x_s^{(2)}\right)^2}{2R_\delta^2}-\frac{\left(y-y_s^{(2)}\right)^2}{2R_\delta^2}-\frac{\left(z-z_s^{(2)}\right)^2}{2R_\delta^2}\right].
\end{align}

\begin{figure}
    \centering
    \includegraphics[width=0.7\textwidth]{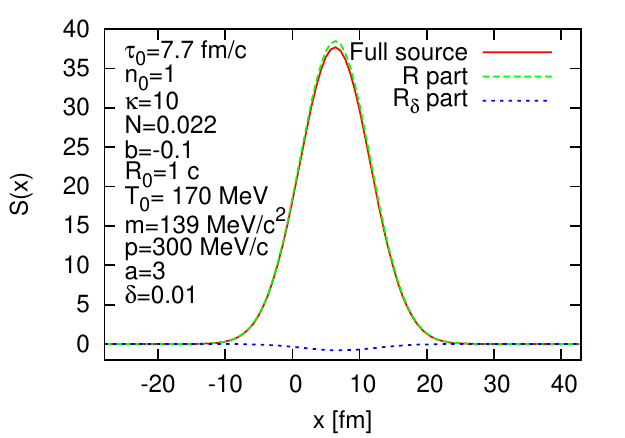}
    \caption{The two component Gaussian source at a given set of parameters denoted on the label.}
    \label{fig:source}
\end{figure}

Here, the $R$ and $R_\delta$ describe the widths of these Gaussian parts of the source. A visualization of the source can be seen in Figure \ref{fig:source}. We can see, that the $\zeta^{(2)}$ term, which has the width $R_\delta$ gives a negative contribution to the source with the chosen set of parameters, however the sign of the perturbative peak depends on the choice of parameters and could yield a positive gain. 

Furthermore, $T_\text{eff}$ and $T_\delta$ are effective temperatures, corresponding to the inverse logarithmic slope of the Maxwell--Boltzmann like distributions. $R$ and $T_\text{eff}$ are the same as in the original Hubble-flow, while $R_\delta$ and $T_\delta$ give the perturbative corrections to the Gaussian width and the effective temperature. The newly introduced notations are the following:

\begin{align}
T_\text{eff}&=T_0+\frac{T_0E\dot{R_0}^2}{2b(T_0-E)}, \hfill &T_\delta=T_0+\frac{T_0E\dot{R_0}^2}{2b(2T_0-E)},\\\label{rhbt}
R^2&=\frac{T_0\tau_0^2(T_\text{eff}-T_0)}{ET_\text{eff}}, \hfill &R_\delta^2=\frac{T_0\tau_0^2(T_\delta-T_0)}{ET_\delta},\\
x_s^{(1)}&=\frac{p_x\tau_0(T_\text{eff}-T_0)}{ET_\text{eff}}, \hfill &x_s^{(2)}=\frac{p_x\tau_0(T_\delta-T_0)}{ET_\delta},\\
y_s^{(1)}&=\frac{p_y\tau_0(T_\text{eff}-T_0)}{ET_\text{eff}}, \hfill &y_s^{(2)}=\frac{p_y\tau_0(T_\delta-T_0)}{ET_\delta},\\
z_s^{(1)}&=\frac{p_z\tau_0(T_\text{eff}-T_0)}{ET_\text{eff}}, \hfill &z_s^{(2)}=\frac{p_z\tau_0(T_\delta-T_0)}{ET_\delta}.
\end{align}

From the source function, the single-particle momentum distribution can be calculated:

\begin{equation}
    N_1(p)=\int \mathrm{d}^4x S(x,p).
\end{equation}

To perform this integral analytically we use the Gaussian saddlepoint approximation. In general, we have the integrand in the form of $f(x)g(x)$, where $f(x)$ is slowly changing, and $g(x)$ has a sharp, unique peak at $x_0$:

\begin{equation}
\int f(x) g(x)=f(x_0)g(x_0)\sqrt{\frac{2\pi}{-(\ln(g(x_0)))''}}.
\end{equation}

From this we can easily get the final form of the single-particle momentum distribution:

\begin{align}
N(p)=Nn_0\mathcal{E}_1\mathcal{V}_1(1+\mathcal{P}_1+\mathcal{P}_2+\mathcal{P}_3)+Nn_0\mathcal{E}_2\mathcal{V}_2(\mathcal{P}_4+\mathcal{P}_5).
\end{align}

Here, we introduced the following functions:

\begin{align}
\mathcal{E}_{1,2}&=\exp\left[-\frac{E^2+m^2}{2ET_0}-\frac{p^2}{2ET_{\text{eff},\delta}}\right],\\
\mathcal{V}_{1,2}&=\sqrt{\frac{2\pi T_0\tau_0^2}{E}\left(1-\frac{T_0}{T_ {\text{eff},\delta}}\right)}^3\left(E-\frac{p^2}{E}\left(1-\frac{T_0}{T_{\text{eff},\delta}}\right)\right).
\end{align}

The terms which come from the first order perturbations are denoted with $\mathcal{P}_i$ and are of the following form in this concrete case of the solution with a saddlepoint approximation:

\begin{align}
    \mathcal{P}_i=\begin{cases}
                   \epsilon_i(x=x_s^{(1)},y=y_s^{(1)},z=z_s^{(1)}), \text{if } i=1,2,3,\\
                   \epsilon_i(x=x_s^{(2)},y=y_s^{(2)},z=z_s^{(2)}), \text{if } i=4,5.
                  \end{cases}
\end{align}

Looking at the final form of the momentum distribution we can see that it is spherically symmetric as we have expected from the spherically symmetric solution.
\section{Discussion}
\begin{figure}
    \centering
    \includegraphics[width=0.7\textwidth]{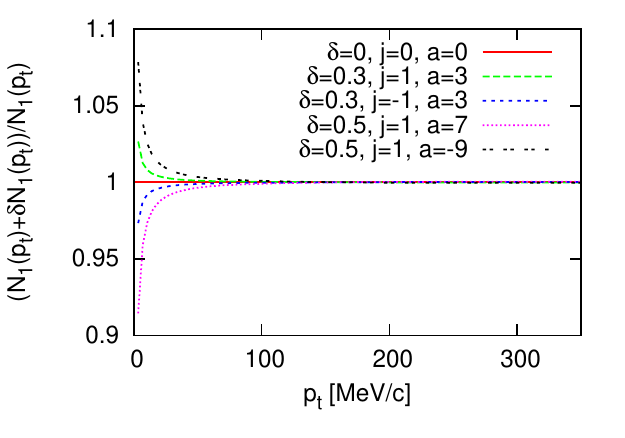}
    \caption{The ratio of the original and the perturbatively corrected single-particle transverse momentum distribution for the investigated solution. The model parameters of the original Hubble-flow come from fits to experimental data \cite{Csanad:2009wc}.}
    \label{fig:n1ratio}
\end{figure}

To understand the role of perturbations on top of the original Hubble-flow we can plot the calculated quantities with given values of parameters. For this we use model parameters of the Hubble-flow from \cite{Csanad:2009wc} where quantities calculated from the exact solution were fitted to the experimental data. With these parameter values we can study the role of acceleration in this concrete solution and the role of the $a$, $\delta$ and $j$ parameters. We can see from Equations \ref{eq:e1} and \ref{eq:e5} that the source and the invariant momentum distribution does not depend separately on $\delta$ or $j$, but on their product $\delta j$. Also, the form of scale parameter does not affect the observables directly, therefore, we can not study the role of these parameters independently: Their product defines the scale of the perturbations. In Figure \ref{fig:n1ratio} we can see the ratio of the original and the perturbated transverse momentum distributions at different values of the $a$ and $\delta j$ parameters with the Gaussian saddlepoint approximation. It can be seen that with this approach, the perturbations only give small corrections to the low momentum region of the single particle momentum distribution. 

However, the saddlepoint approximation might not give back all the properties of the perturbation, as it assumes that the function that multiplies the Gaussian peak is slowly changing. In our case we can see from Equations \ref{eq:e1} and \ref{eq:e5} that we have terms proportional to $\tau_0/r$ that might influence the result, as $r/\tau_0\ll 1$. Therefore we could make a Laurent-expansion of the terms $\epsilon_i$; as it turns out the series is finite in the negative region with all the terms vanishing below $(r/\tau_0)^{-2}$, which indicates that all the terms are integrable. This approach gives rise to rather complicated integrals and we will not discuss this method further, we simply wanted to note the possibility of such a calculation in the future. For this type of calculation, it is however sufficient to use the saddlepoint calculation, as it provides a good approximation of the results if the requirement $T_0/T_{\text{eff},\delta}\approx 1$ is met, but $p/E\ll1$ is not.

Let us now turn to study the geometry of the particle emitting source. From femtoscopic measurements, the homogeneity region of the source can be mapped out. The first intensity correlation measuruments were carried out by R. Hanbury Brown and R. Q. Twiss, thus these are often called HBT measurements.\cite{HanburyBrown:1956bqd} The size of the source can be characterized by the HBT-radii, which are often associated with the Gaussian widths of the source \cite{Adler:2004rq,Afanasiev:2009ii}. However, let us note here that there are more general approaches to characterize the source \cite{Adare:2017vig,Csorgo:2003uv}. In this paper, we have used a Gaussian approximation for the analytic calculations, therefore we can associate the Gaussian width of the source with the HBT-radius of the studied, spherically symmetric system. The source is the sum of two terms with different widths. This gives us two different HBT-radii, $R$ and $R_\delta$, where $R$ is the same as it is for the exact solution \cite{Csanad:2009wc}. The HBT-radius of such a source is some average of the radii $R$ and $R_\delta$.

\begin{figure}
    \centering
    \includegraphics[width=0.7\textwidth]{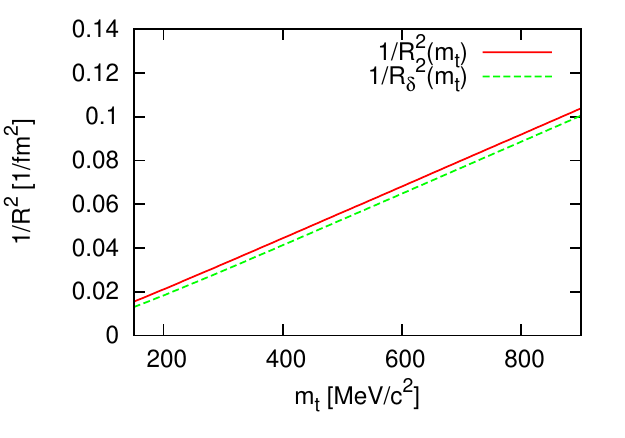}
    \caption{We can see the transverse mass scaling of the calculated HBT-radii, which is usually observed in experimental data.}
    \label{fig:hbt2}
\end{figure}

The values of $R$ and $R_\delta$ do not depend on the perturbation parameters $\delta$, $j$ and $a$, but their averaging does depend on the choice of these. For such model parameters as used for Figure  \ref{fig:n1ratio} the average HBT-radius is approximately the same as the original $R$, and only for large $\delta$ and $a$ values do we get a significant contribution from $R_\delta$.  We can look at the HBT-radius as the function of the transverse mass: $m_\text{t}=\sqrt{m^2+p_\text{t}^2}$. Experimentally the HBT-radii usually show a scaling, regardless of particle species, collision energy or centrality \cite{Adler:2004rq,Afanasiev:2009ii}. The cause of this scaling is the hydrodynamical expansion both in the longitudinal and the radial directions \cite{Csorgo:1995bi}. We can see the $R\propto 1/\sqrt{m_\text{T}}$ scaling in Figure \ref{fig:hbt2} as it was already shown in \cite{Kurgyis:2018sgx}.

\section{Summary}
We have given the perturbated source function for the perturbative, accelerating generalization of the exact Hubble-flow, and calculated the single-particle momentum distribution and the HBT-radius for a spherically symmetric solution. This way the solution includes the acceleration and pressure gradient. For the observables we have found that the perturbations cause only small deviations from the original quantities in the Gaussian saddlepoint approximation. Also, we have seen that the source is a sum of two Gaussians with different widths. Furthermore, we have found that the choice of scale parameter does not affect the calculated observables directly, but results only in a difference in the perturbation scale. For further studies, the elliptical flow could be also calculated, but in a non-spherically symmetric case.

\end{document}